\documentclass[letterpaper]{article}
\usepackage{natbib,alifeconf}
\usepackage{topcapt}
\usepackage{booktabs}
\newcommand{\be}{\begin{eqnarray}}
\newcommand{\ee}{\end{eqnarray}}

\title{Does self-replication imply evolvability?} 
\author{Thomas LaBar$^{1,2}$, Christoph Adami$^{1,2}$ \and Arend Hintze$^{2,3,4}$ \\
\mbox{}\\
$^1$Department of Microbiology and Molecular Genetics \\
$^2$BEACON Center for the Study of Evolution in Action \\
$^3$Department of Integrative Biology\\
$^4$Department of Computer Science and Engineering\\ Michigan State University, East Lansing, MI 48824\\
hintze@msu.edu}

\begin{document}
\maketitle

\begin{abstract}
The most prominent property of life on Earth is its ability to evolve. It is often taken for granted that self-replication--the characteristic that makes life possible--implies evolvability, but many examples such as the lack of evolvability in computer viruses seem to challenge this view. Is evolvability itself a property that needs to evolve, or is it automatically present within any chemistry that supports sequences that can evolve in principle? Here, we study evolvability in the digital life system Avida, where self-replicating sequences written by hand are used to seed evolutionary experiments. We use 170 self-replicators that we found in a search through 3 billion randomly generated sequences (at three different sequence lengths) to study the evolvability of {\em generic} rather than hand-designed 
self-replicators. We find that most can evolve but some are evolutionarily sterile. From this limited data set we are led to conclude that evolvability is a likely--but not a guaranteed-- property of random replicators in a digital chemistry.


\end{abstract}

\section{Introduction}
For life of the type as we experience it on Earth to emerge from an initially abiotic state requires two seemingly independent things to happen. First, a self-replicator has to emerge (or else, to arrive on Earth from extraterrestrial sources~\citep{Arrhenius1908,HoyleWickramasinghe1981,Wickramasinghe2011}). Secondly, this self-replicator must be able to evolve and thus diversify into the complexity we see today. In general we think of self-replicators as extremely rare~\citep{vonNeumann1966}. But in order to jump-start the process of evolution, these rare self-replicators now also must be endowed with another property--evolvability~\citep{WagnerAltenberg1996,Kirschner1998,Wagner2005}. By evolvability here we mean the ability to produce variants or mutants (produced by a faulty self replication process or else by external noise) that can also self-replicate. Without evolvability the self-replicator would just multiply but not adapt. While it is safe to assume that one of the first self-replicators on Earth was evolvable ($n=1$), it is not at all clear whether evolvability is a general property of self-replicators, or else is an additional constraint that renders the emergence of life even more improbable. Here we use the computational evolution system Avida~\citep{AdamiBrown1994,Adami1998,Ofriaetal2009} to investigate whether random self-replicators, that is, randomly generated sequences of code written in the avidian instruction set that happen to be able to self-replicate, also are automatically endowed with the capacity to evolve. Even though evolvability appears to be inherent to avidians, we cannot rule out a priori whether the observed evolvability of avidians is a consequence of the evolvabilty of the hand-written ancestor or is instead a germane property of all self-replicators that can exist within this digital chemistry.  To resolve this question, we searched for self-replicating sequences that we generated randomly, within the three size classes $L=8$, $L=15$ (the size of the standard avidian self-replicator), and $L=30$. We found 170 such sequences after generating 3 billion random sequences, and report their evolvabilty below.

\section{Self-replicators in Avida}
\begin{figure}[!htbp] 
   \centering
   \includegraphics[width=3.5in]{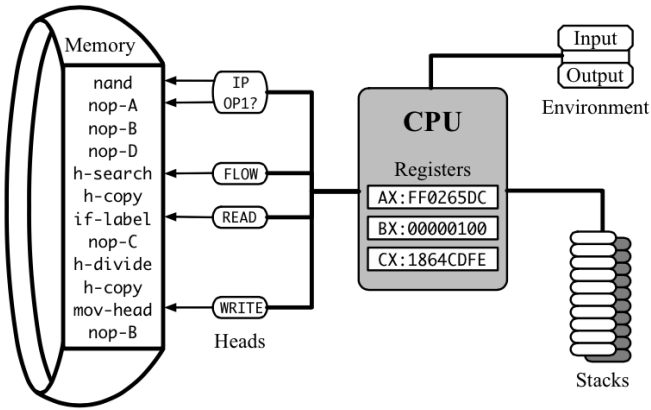} 
   \caption{The avidian CPU in the process of executing a segment of code. The CPU uses three registers (AX,BX,CX) as well as an instruction pointer (IP) that reads the program into the CPU.  A read-head, a write-head, and a flow-head are used to specify positions in the CPU's memory. The `copy' command reads from the read-head and writes to the write-head, while `jump'-type statements move the instruction pointer to the flow-head. The CPU uses two stacks to simulate an ``infinite Turing tape", while input/output buffers serve to communicate between the CPU and its environment (reproduced from~\cite{Ofriaetal2009}, with permission). }
   \label{avida}
\end{figure}

In Avida, small self-replicating computer programs (``avidians") compete for limited memory space and limited CPU resources needed to successfully self-replicate (see \citealt{Ofriaetal2009} for a complete description of the Avida system). This ability to self-replicate is contained within an individual avidian's genome of instructions (see Fig.~\ref{avida}). Because this genome is then passed on to an avidian's descendants, the ability to self-replicate is heritable. Selection in Avida is implicit, since faster replicators have a higher chance of making offspring.  During the process of self-replication, mutations may be introduced, resulting in error-prone replication and variation within the population. Thus, Avida is an {\em instance} of evolution by natural selection~\citep{Pennock2007}, because it contains inheritance, variation, and differential fitness among individuals. Avida has been used to explore a diverse set of topics in evolutionary biology such as the evolution of genomic complexity~\citep{lenski1999genome}, the ``survival of the flattest" effect at high mutation rates~\citep{wilke2001evolution}, the evolution of complex features~\citep{lenski2003evolutionary}, the evolution of reproductive division of labor~\citep{goldsby2014evolutionary}, and the role of coevolution in the origin of complexity~\citep{zaman2014coevolution}.

Avidian genomes are composed of usually 26 different instructions, typically rendered as a string of lowercase letters, where each letter corresponds to one command (see Table 1). These instructions can be considered analogous to the 20 amino acids common to all biological organisms. 
\begin{table*}[htbp]
   \centering
   \begin{tabular}{@{} lll @{}} 
      \toprule
      Instruction    & Description & Symbol\\
      \midrule
nop-A    & no operation (type A) & a \\
nop-B   & no operation (type B) & b \\
nop-C   & no operation (type C) & c \\
if-n-equ & Execute next instruction only-if ?BX? does not equal complement & d\\
if-less & Execute next instruction only if ?BX? is less than its complement & e\\
if-label & Execute next instruction only if template complement was just copied & f\\
mov-head & Move instruction pointer to same position as flow-head & g\\
jmp-head & Move instruction pointer by fixed amount found in register CX & h\\
get-head & Write position of instruction pointer into register CX & i\\
set-flow & Move the flow-head to the memory position specified by ?CX? & j\\
shift-r & Shift all the bits in ?BX? one to the right & k\\
shift-l & Shift all the bits in ?BX? one to the left & l\\
inc & Increment ?BX? & m\\
dec & Decrement ?BX? & n\\
push &       Copy value of ?BX? onto top of  current stack & o\\
pop & Remove number from current stack and place in ?BX? & p\\
swap-stk & Toggle the active stack & q\\
swap & Swap the contents of ?BX? with its complement & r\\
add & Calculate  sum of BX and CX; put  result in ?BX? & s\\
sub & Calculate  BX minus CX; put result in ?BX? & t\\
nand & Perform bitwise NAND on BX and CX; put  result in ?BX? & u\\
h-copy &  Copy instruction from read-head to write-head and advance both & v\\
h-alloc & Allocate memory for offspring & w\\
h-divide & Divide off an offspring located between read-head and write-head & x \\
IO &  Output value ?BX? and replace with new input & y\\
h-search & Find complement template and place flow-head after it & z\\
      \bottomrule
   \end{tabular}
   \caption{Instruction set of the avidian programming language used in this study. The notation ?BX? implies that the command operates on a register specified by the subsequent nop instruction (for example, nop-A specifies the AX register, and so forth). If no nop instruction follows, use the register BX as a default. More details about this instruction set can be found in~\cite{Ofriaetal2009}.}
   \label{avidainst}
\end{table*}

A self-replicator in Avida must have a set of instructions in the right order to: first allocate memory, then copy itself into this newly allocated memory, and finally it has to split the new memory from the old one in order to crate a new organism. These instructions and their interactions are complex and in general it is not possible to predict if an organism will replicate just by examining the sequence. As a consequence, the ability to self-replicate has to be tested directly by allowing the instructions to execute (we surmise that this problem of predicting the ability to self-replicate is similar to the Halting Problem~\citep{turing1936computable}, where it is necessary to run the algorithm to find out if it halts or not).

In the majority of all Avida experiments a default hand-designed ``start" organism is used to seed a population. This choice is to some extent historical: the designers assumed that self-replicators were too rare to be found by a random process~\citep{Adami1998}. Indeed, any particular sequence of length $L$=15 for example (the length of one of the standard hand-written ancestors) has a likelihood of $p=26^{-15}\approx 6\times 10^{-22}$. If a million of such sequences could be checked per second, on one thousand CPUs running in parallel, it would take about 50,000 years to find it. As a remedy, the designers wrote one instead (most of the common ancestors in Avida were written by Charles Ofria). However, it is clear that the density of self-replicators in the space of all sequences depends on the chemistry (here, the instruction set) used. Since the inception of Avida in 1993, the standard instruction set has changed, and it appears that using the current set (that is, the current ``chemistry") self-replicators can be found among randomly generated sequences (as reported here) because the information content of the sequences is significantly smaller than 15.

No self-replicator in nature (or within Avida) replicates without error, because noise in the system is inevitable and will always affect replication by increasing variation. Variation in Avida can have two different causes. First, it  is possible that the sequence of commands performs in such ways that the resulting copy is modified. This typically leads to repeated commands, insertions, or deletions. Such changes are inherent to the self-replicator (they are genetically encoded and thus deterministic) and referred to as ``implicit" mutations. A second mechanism is less deterministic, and occurs during the processes of division and reproduction, processes that are inherently made to be error-prone to a degree that can be specified by the user. These copy-errors, as well as the probabilistic insertion and deletion of instructions or code snippets reflect the noisiness of the system and are not under control of the organism itself. Note that even the most sophisticated polymerases that also have proof reading abilities cannot create perfect replicates every time, while this is in principle possible for avidians without implicit mutations by turning the mutation rate to zero. While biochemical-based life does not have a mutational mechanism similar to the implicit mutations seen in avidians, it is not unreasonable to assume that early replicators also underwent large mutations due to a lack of error-correction during replication.

\section{Evolvability in Avida}
Evolvability describes the ability of an organism to undergo evolutionary adaptation~\citep{WagnerAltenberg1996,Kirschner1998,Wagner2005}. There are two main pathways to adaptation in Avida. The first is {\em optimization}, where organisms with a minimal replication time outcompete others in the population (akin to $r$-selection~\citep{Pianka1970}). The second is {\em innovation}, where avidians can evolve new phenotypic traits that enable a fitness increase (the $K$-selection mode). These phenotypic traits are the ability to perform certain Boolean logic operations; these logic operations allow an individual to execute a greater proportion of its genome than other avidians that do not perform such logic operations. This indirectly allows them to self-replicate at a greater rate (see, for example,~\citealt{Adami1998,Adami2006a}). These different modes of survival can be thought of as different niches that avidians can inhabit~\citep{WhiteAdami2004}.

It is possible in general that an organism (digital or otherwise) carries such a specific sequence of nucleotides or (as in Avida) commands that every possible mutation prevents self-replication. From a fitness landscape point of view, such replicators would represent an isolated fitness peak where self replication is only possible at the top, and where each mutation is lethal. Such self-replicators, born at the top of the fitness peak, so to speak, would be unevolvable. A self-replicator that can evolve, on the other hand,  would never emerge at peak in the landscape, but rather on a lower level and subsequently find positive and or neutral mutations that ultimately lead to higher fitness levels in the landscape. We know that the hand-written replicators in Avida are evolvable, but it is not clear how likely evolvability is among randomly generated replicators.

\section{Methods}
All experiments are performed using Avida version 2.14, which can be downloaded from https://github.com/devosoft/avida. We first randomly generated Avida sequences to discover individuals that could self-replicate, using a uniform random distribution for each command at each site, which ensures that each sequence has the same likelihood to be generated, given by
\be{}
p=\frac{1}{D^{L}}\;,
\ee{} 
where $D$ is the size of the alphabet ($D$=26 here) and $L$ is the length of the sequence generated~\citep{AdamiLaBar2015}. To decide whether a sequence could successfully self-replicate, it must pass two tests. First, we tested whether the organism could successfully divide within its lifespan. Here, we used standard  Avida parameters for an organism's lifespan: it must divide before it executes $20\times L$ instructions. This test indicates that an avidian can successfully reproduce, it does not imply that its descendants also can reproduce. In our search we discovered many viable avidians that were able to successfully divide into two non-viable organisms. Therefore, we only counted sequences that could replicate and produce offspring that could also replicate as true self-replicators (in other words, they are ``colony-forming"). This does not imply that every self-replicator produces a perfect copy of itself in the absence of mutation. Indeed, most of these replicators undergo implicit mutations solely due to their genome sequence, and their offspring differ in length from the parent. In analyzing a genome's ability to self-replicate, we used the default Avida settings, described for example in~\citep{Ofriaetal2009}. 
For our study of evolvability, we also included the default length 15 hand-written ancestor in our set of self-replicators. 

In order to test whether these self-replicators could optimize their fitness, we evolved them in an environment where {\em only} decreased self-replication speed was under positive selection (the $r$-selection niche). We evolved these replicators for $10^3$ generations at a population size of $3,600$ individuals (10 replicates). Instruction mutations occurred at a genomic rate of 0.1 at division (meaning the likelihood to incur an error is length-independent), and both insertions and deletions occurred at a genomic rate of 0.005 per division. We measured an organism's evolvability as its gain in relative fitness at the end of the experiment. 

Finally, we tested the self-replicators' ability to evolve greater complexity by evolving new phenotypic traits (that is, in the $K$-selection mode). The traits that are rewarded are logical functions that an avidian can solve by stringing together code involving the {\tt nand} function (given by the letter {\tt u}, see Table 1). In this setting, nine different logic tasks may be rewarded, denoted as the bit-wise {\tt NOT, NAND, AND, OR-NOT,OR, AND-NOT, NOR, XOR}, and {\tt EQU}. We evolved the self-replicators of length 15 in an environment where nine Boolean logic operations, and hence nine phenotypic traits, are under positive selection; this is often referred to as the logic-9 environment~\citep{lenski2003evolutionary}. For this experiment, we evolved these replicators for $10^4$ generations at a population size of $10^4$ individual (ten replicates). Increased population size and experiment time was used to allow time for trait evolution. The mutation rates were the same as in the optimization experiments. We quantified an organism's evolvability in these experiments by measuring the number of evolved phenotypic traits.

\section{Results}
\begin{figure}[htb]
\begin{center}
\includegraphics[width=8cm]{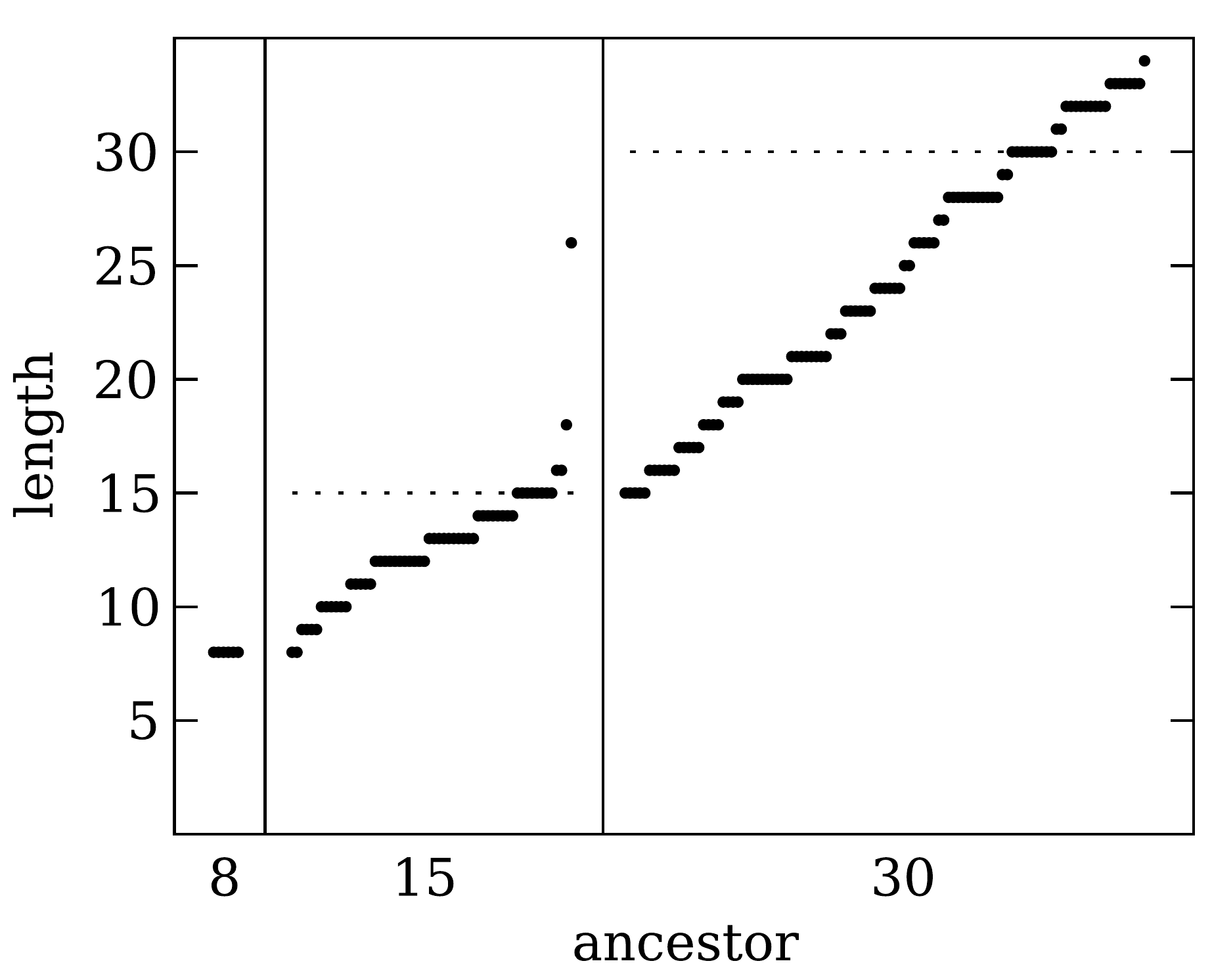}
\caption{Change in length after the first round of replication for all self replicators we found of length 8, 15, and 30. The self-replicators are ordered by length increase within their individual groups. The dashed lines indicate the start organism's length.}
\label{fidelityAll}
\end{center}
\end{figure}

Of the $10^9$ randomly-generated Avidian genomes in each length class, we found 6 self-replicators of length 8, 58 self-replicators of length 15, and 106 self-replicators of length 30 (see Table~\ref{tab:seq} for their sequences). However, it is unlikely that each self-replicator needs every instruction in its genome in order to self-replicate~\citep{Adami2015}. Many of these self-replicators could be unstable and would generate implicit mutations upon replication. Therefore, we tested the replication fidelity of each self-replicator (Fig.~\ref{fidelityAll}). All replicators of length 8 were able to undergo error-free replication without external noise (mutation rate set to zero). This is likely due to the fact that avidian genomes cannot be shorter than length 8; however, that does not prevent the occurrence of genomes that replicate to a length larger than 8, which we did not see among this set. Self-replicators of length 15 and 30 could not replicate error-free, even without external noise, and most of their genome sequences decreased upon successful replication. 

\begin{figure}[htb]
\begin{center}
\includegraphics{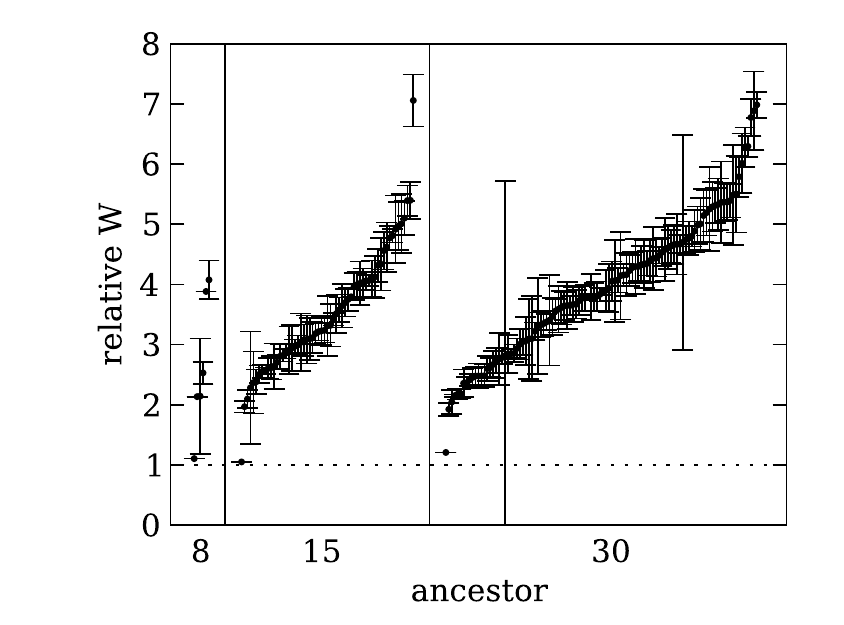}
\caption{Relative fitness after 1,000 generatios of evolution for all replicators we found of length 8, 15, and 30. The replicators are ordered by relative fitness increase within their groups of similar length. The fitness of all start organisms is normalized to $1$, indicated by the dashed line. The error bars indicate the standard deviation over $10$ replicate evolution trials with the same ancestor.}\label{relativeFitnessAll}
\end{center}
\end{figure}

Next, we tested the evolvability of these replicators in the sense of how well they could  optimize their genomes as a way to increase replication speed, and thus their fitness. We found that the majority of these spontaneous self-replicators of length 8, 15, and 30 possess the ability to optimize their replication algorithm (see Figure~\ref{relativeFitnessAll}). While the increase in fitness across the different self-replicators varied due to differing ancestral fitness, most increased in fitness by more than two-fold over the course of $10^3$ generations of evolution. However, we found a few replicators (~3\%) that were evolutionary sterile (defined as a relative fitness $<2$), demonstrating that not all self-replicators can easily undergo significant adaptation.

\begin{figure}[htb]
\begin{center}
\includegraphics[width=3.5in]{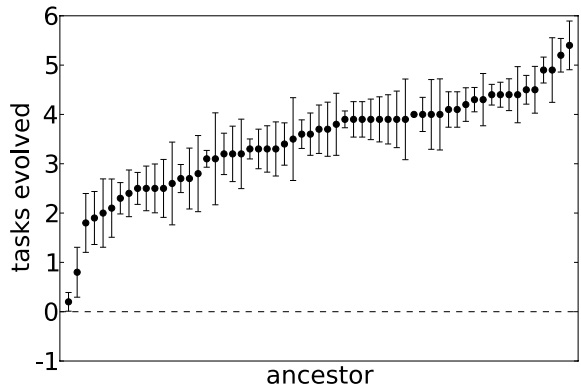}
\caption{The number of logic tasks that spontaneous replicators of length 15 were able to perform after $10,000$ generations of evolution in the Avida logic-9 environment. Replicators are ordered by number of tasks achieved. Error bars indicate the standard error of the mean for 10 replicates per ancestral organism.}
\label{traitsLength15}
\end{center}
\end{figure}

Because the ability to optimize replication speed does not automatically imply the ability to evolve greater complexity, we also tested at the ability of the length 15 replicators to evolve new phenotypic traits (see Methods for a description of those traits). All 58 self-replicators were able to evolve at least one phenotypic trait in one replicate, although the likelihood of the emergence of phenotypic traits varied greatly across the different replicators and within the replicates for each specific replicator (see Fig.~\ref{traitsLength15}). Moreover, most self-replicators were able to evolve multiple traits. The hand-written Avida ancestor of length 15 displayed the {\em least} evolvability in regards to trait evolution compared to the 58 randomly-generated self-replicators. Only one out of ten replicates evolved any phenotypic traits in the allotted time, although that replicate did manage the evolution of two traits.

\section{Discussion}
Here, we asked if spontaneous self-replicators in Avida would all possess the additional ability to change in such a way that they could initiate evolution. We showed that the majority of self-replicators are robust enough to tolerate mutations as well as changes to their genome size. This suggests that self-replicators in this digital chemistry are robust and thus will most likely be able to jump-start evolution. Further, this result holds both when we looked at the ability of self-replicators to increase fitness through optimization or through the evolution of new phenotypes. In addition, many spontaneous self-replicators made faulty copies of themselves in the absence of external noise, which further supports the idea that variability is an inherent property of spontaneous self-replication.

\begin{figure}[htb]
\begin{center}
\includegraphics[width=3.5in]{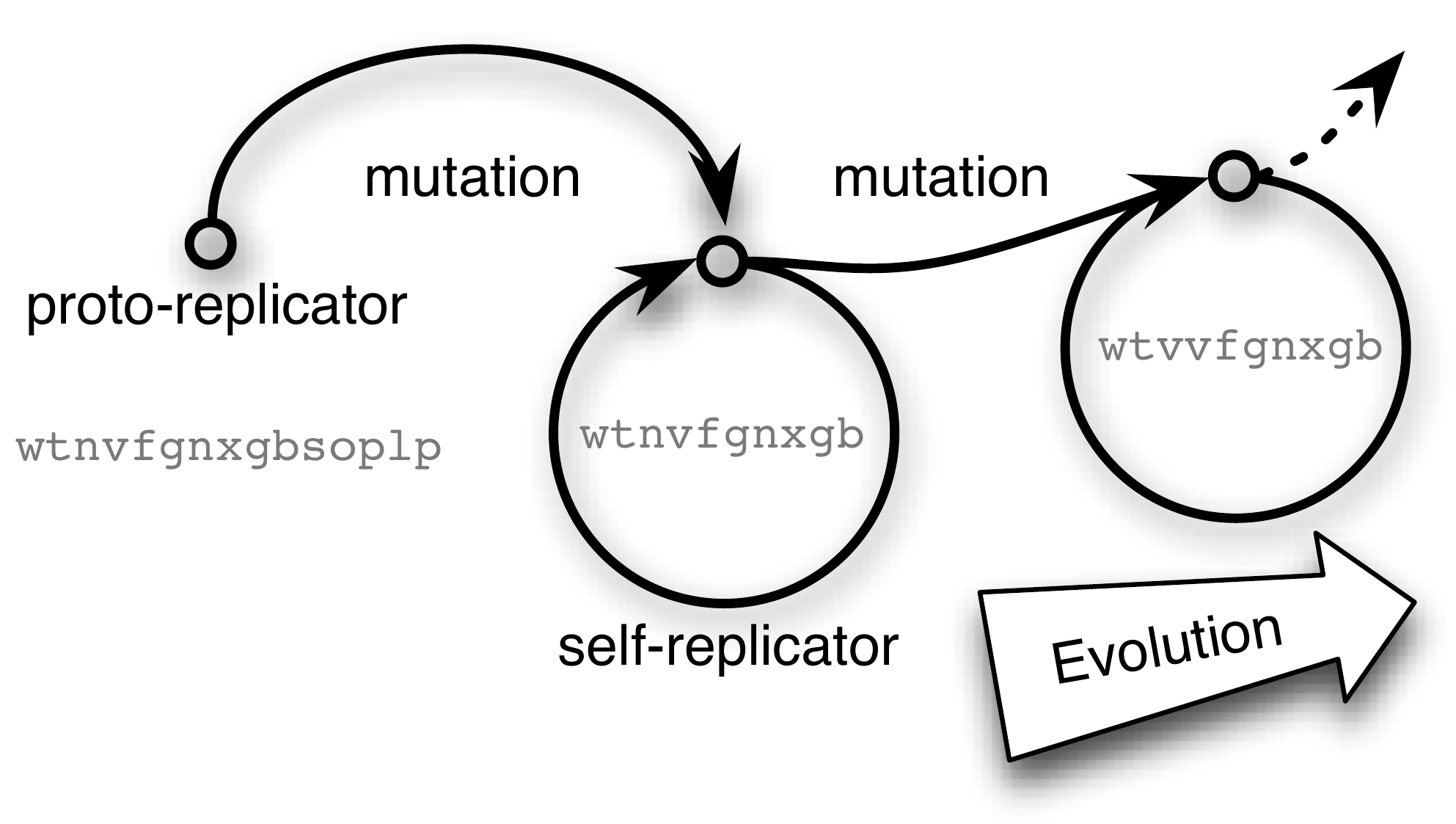}
\caption{Illustration how an Avida proto-replicator (left) ``copies" itself into a self-replicator (middle). This replicator can now experience external mutations, which leads to a new self replicator. The repetition of such process is what jump-starts evolution (right). The text string is a program example from Avida.}
\label{overview}
\end{center}
\end{figure}

These results confirm that not every spontaneous self-replicator necessarily is evolvable. On the other hand, we also find that the majority of spontaneous self-replicators deterministically do not make exact copies even in the absence of mutation, questioning the term ``self-replicator" for these genotypes. Indeed, while these spontaneously generated sequences are ``colony-forming" (in the sense that they produce sequences that can also replicate themselves), these are not self-replicators in the strictest sense of the definition. Instead, these replicators mutate themselves to become a self-replicator: we call these individuals ``proto-self-replicators" or ``proto-replicators". This means that the number of potential sequences that can start evolution is increased by existence of these proto-replicators. In natural chemistries, this phenomenon has been discussed widely~\citep{bernal1949physical,miller1953production,eigen1971,miller1974origins,eschenmoser1992chemistry}, suggesting that the first self-replicator might have needed either self organisation or some other forms of catalysis or autocatalysis in order to create the right chemistry to start polymerization in the first place. Figure \ref{overview} gives an illustration of this concept. 

Leo Tolstoy famously begun his ``Anna Karenina" by remarking that {\em``Happy families are all alike; every unhappy family is unhappy in its own way."}. In the same manner we asked here whether, with respect to evolvability, all self-replicators are alike (while of course all non-replicators are non-replicating in their own way). Our results show that not every self-replicator is suitable as the ancestor for evolution. But we also find that, with respect to the capacity to evolve functional complexity, random replicators are better than even those designed by thinking humans.

Our results further suggest that life, if possibly found elsewhere, does not necessarily experience evolution. We might at some point in the future find self-replicators that maintained their ancestral form due to their inability to tolerate enough variation to evolve. Similarly, it seems possible that, if such an mutationally-fragile self-replicator exists, it might be outcompeted by other self replicators that can take advantage of mutations, and thus attain higher fitness and ultimately outcompete those  mutationally-fragile self-replicators. Thus, as long as evolvability is evolvable, complexity should ensue even when sparked by the most humble and awkward replicators. 

\footnotesize

\begin{thebibliography}{}

\bibitem[Adami, 1998]{Adami1998}
Adami, C. (1998).
\newblock {\em Introduction to Artificial Life}.
\newblock Springer Verlag, New York and Heidelberg.

\bibitem[Adami, 2006]{Adami2006a}
Adami, C. (2006).
\newblock Digital genetics: Unravelling the genetic basis of evolution.
\newblock {\em Nat Rev Genet}, 7:109--118.

\bibitem[Adami, 2015]{Adami2015}
Adami, C. (2015).
\newblock Information-theoretic considerations concerning the origin of life.
\newblock {\em Origins of Life and Evolution of Biospheres}, 45.

\bibitem[Adami and Brown, 1994]{AdamiBrown1994}
Adami, C. and Brown, C. (1994).
\newblock Evolutionary learning in the {2D} {Artificial Life} system {Avida}.
\newblock In Brooks, R. and Maes, P., editors, {\em Proc. Artificial Life 4},
  pages 377--381. MIT Press.

\bibitem[Adami and LaBar, 2015]{AdamiLaBar2015}
Adami, C. and LaBar, T. (2015).
\newblock From entropy to information: Biased typewriters and the origin of
  life.
\newblock In Walker, S., Davies, P., and Ellis, G., editors, {\em Information
  and Causality: From Matter to Life}. Cambridge University Press, Cambridge,
  MA.

\bibitem[Arrhenius, 1908]{Arrhenius1908}
Arrhenius, S. (1908).
\newblock {\em Worlds in the Making: The Evolution of the Universe}.
\newblock Harper and Row, New York, NY.

\bibitem[Bernal, 1949]{bernal1949physical}
Bernal, J. (1949).
\newblock The physical basis of life.
\newblock {\em Proceedings of the Physical Society. Section B}, 62:597.

\bibitem[Eigen, 1971]{eigen1971}
Eigen, M. (1971).
\newblock Selforganization of matter and the evolution of biological
  macromolecules.
\newblock {\em Naturwissenschaften}, 58:465--523.

\bibitem[Eschenmoser and Loewenthal, 1992]{eschenmoser1992chemistry}
Eschenmoser, A. and Loewenthal, E. (1992).
\newblock Chemistry of potentially prebiological natural products.
\newblock {\em Chemical Society Reviews}, 21:1--16.

\bibitem[Goldsby et~al., 2014]{goldsby2014evolutionary}
Goldsby, H.~J., Knoester, D.~B., Ofria, C., and Kerr, B. (2014).
\newblock The evolutionary origin of somatic cells under the dirty work
  hypothesis.
\newblock {\em PLoS biology}, 12:e1001858.

\bibitem[Hoyle and Wickramasinghe, 1981]{HoyleWickramasinghe1981}
Hoyle, F. and Wickramasinghe, N. (1981).
\newblock {\em Evolution from Space}.
\newblock Simon \& Schuster Inc, New York, NY.

\bibitem[Kirschner and Gerhart, 1998]{Kirschner1998}
Kirschner, M. and Gerhart, J. (1998).
\newblock Evolvability.
\newblock {\em Proc. Natl. Acad. Sci. U.S.A.}, 95:8420--8427.

\bibitem[Lenski et~al., 1999]{lenski1999genome}
Lenski, R.~E., Ofria, C., Collier, T.~C., and Adami, C. (1999).
\newblock Genome complexity, robustness and genetic interactions in digital
  organisms.
\newblock {\em Nature}, 400:661--664.

\bibitem[Lenski et~al., 2003]{lenski2003evolutionary}
Lenski, R.~E., Ofria, C., Pennock, R.~T., and Adami, C. (2003).
\newblock The evolutionary origin of complex features.
\newblock {\em Nature}, 423:139--144.

\bibitem[Miller et~al., 1953]{miller1953production}
Miller, S.~L. et~al. (1953).
\newblock A production of amino acids under possible primitive earth
  conditions.
\newblock {\em Science}, 117:528--529.

\bibitem[Miller and Orgel, 1974]{miller1974origins}
Miller, S.~L. and Orgel, L.~E. (1974).
\newblock {\em The origins of life on the earth}.
\newblock Prentice-Hall Englewood Cliffs, NJ.

\bibitem[Ofria et~al., 2009]{Ofriaetal2009}
Ofria, C., Bryson, D., and Wilke, C. (2009).
\newblock Avida: A software platform for research in computational evolutionary
  biology.
\newblock In Adamatzky, A. and Komosinski, M., editors, {\em Artificial Life
  Models in Software}, pages 3--35. Springer Verlag, 2nd edition.

\bibitem[Pennock, 2007]{Pennock2007}
Pennock, R.~T. (2007).
\newblock Models, simulations, instantiations and evidence: The case of digital
  evolution.
\newblock {\em Journal of Experimental and Theoretical Artificial
  Intelligence}, 19:29--42.

\bibitem[Pianka, 1970]{Pianka1970}
Pianka, E. (1970).
\newblock On r and {K} selection.
\newblock {\em American Naturalist}, 104:592--597.

\bibitem[Turing, 1936]{turing1936computable}
Turing, A.~M. (1937).
\newblock On computable numbers, with an application to the
  {E}ntscheidungsproblem.
\newblock {\em Proc. London Math. Soc. Ser. 2}, 42:230-265.

\bibitem[von Neumann, 1966]{vonNeumann1966}
von Neumann, J. (1966).
\newblock {\em The Theory of Self-reproducing Automata}.
\newblock edited and completed by {A. Burks}. Univ. of Illinois Press, Urbana,
  IL.

\bibitem[Wagner, 2005]{Wagner2005}
Wagner, A. (2005).
\newblock {\em Robustness and Evolvability in Living systems}.
\newblock Princeton University Press, Princeton, NJ.

\bibitem[Wagner and Altenberg, 1996]{WagnerAltenberg1996}
Wagner, G.~P. and Altenberg, L. (1996).
\newblock Complex adaptations and the evolution of evolvability.
\newblock {\em Evolution}, 50:967--976.

\bibitem[White and Adami, 2004]{WhiteAdami2004}
White, J.~S. and Adami, C. (2004).
\newblock Bifurcation into functional niches in adaptation.
\newblock {\em Artif Life}, 10:135--44.

\bibitem[Wickramasinghe, 2011]{Wickramasinghe2011}
Wickramasinghe, C. (2011).
\newblock Bacterial morphologies supporting cometary panspermia: A reappraisal.
\newblock {\em International Journal of Astrobiology}, 10:25--30.

\bibitem[Wilke et~al., 2001]{wilke2001evolution}
Wilke, C.~O., Wang, J.~L., Ofria, C., Lenski, R.~E., and Adami, C. (2001).
\newblock Evolution of digital organisms at high mutation rates leads to
  survival of the flattest.
\newblock {\em Nature}, 412:331--333.

\bibitem[Zaman et~al., 2014]{zaman2014coevolution}
Zaman, L., Meyer, J.~R., Devangam, S., Bryson, D.~M., Lenski, R.~E., and Ofria,
  C. (2014).
\newblock Coevolution drives the emergence of complex traits and promotes
  evolvability.
\newblock {\em PLoS Biology}, 12:e1002023.

\end{thebibliography}

\clearpage

\onecolumn
\section{Appendix: Sequences of replicators}
\vskip 2cm
\noindent Table 2: Sequences of 8-mer replicators and 15-mer replicators. The sequence denoted (1) is the hand-written $L$=15 ancestor. The 106 proto- and self-replicators of length $L$=30 can be downloaded from http://dx.doi.org/10.6084/m9.figshare.1416247
\vskip 0.5cm
\centering
   \begin{tabular}{ llll } 
      \toprule
                  {$L=8$} \\
{\tt qxrchcwv}&
{\tt vxfgwjgb}&
{\tt wxvxfggb}&
{\tt vhfgxwgb}\\
{\tt wxrchcvz}&
{\tt wvfgjxgb}\\
$L=15$\\
{\tt wtnvfgnxgbsoplp}&
{\tt mwtepvfgskvrxgb}&
{\tt wvxvfagbchcmwse}&
{\tt nvlvfgqnsxwgbzf}\\
{\tt wlmvmfgowxdogbf}&
{\tt wvfgmwxgbjqpylp}&
{\tt vovvwnfgxlrgbjn}&
{\tt prvfgemsxwepgbo}\\
{\tt vfgxqhmwmfjphgb}&
{\tt vtwfgxgbyhdnahk}&
{\tt vlfgvuiofxwgbpc}&
{\tt dywqvphfguxqdgb}\\
{\tt wvufguuxltsgbwd}&
{\tt vmfifgwvpowxgbt}&
{\tt wzvfgqxrpoujgbn}&
{\tt jivfgzwkjxogbtw}\\
{\tt wvfgnxwhgbaplye}&
{\tt vwowfgqxqwxgbjo}&
{\tt vvfgwxqwdfogbhq}&
{\tt wvfgofeoxxgbrmg}\\
{\tt wvfguzxqjgbiokn}&
{\tt vkhwfgyfrxwgbtr}&
{\tt vfgwvtljvrwxgbl}&
{\tt vfglqwxljgbwdsf}\\
{\tt lwvfgxoetfdgbhp}&
{\tt lsvrwfgxmmwgbwg}&
{\tt vnfgudsftwxwhgb}&
{\tt vlfgvmhwuxwlgbq}\\
{\tt rrowvmfgxjgbuyx}&
{\tt drvfgwioxrmgbjx}&
{\tt inoidjwvfglxgbd}&
{\tt iwlkvfgxgbsslez}\\
{\tt vfgwpxpgbyxxddi}&
{\tt nvqwqfgfnoxpgbm}&
{\tt vfgkqldxidwgbxt}&
{\tt vnwdfgxgbyeevbg}\\
{\tt vmfgqwrmdkyxuhc}&
{\tt qvufgxwdgbojyom}&
{\tt vhfgtxlwgbfbryb}&
{\tt wrvvnfgxmgbctol}\\
{\tt vwqqfgmxgbeixsh}&
{\tt vqvfgvqxwygbwzn}&
{\tt vyhfgxwkypgbyny}&
{\tt yvdwfgxgbvwrfpg}\\
{\tt wvfghqtzoxjirgb}&
{\tt irwovfgxjgbwhbr}&
{\tt wvfgxdmoprllwgb}&
{\tt liwvlfguxgbnhsn}\\
{\tt vfgfyxnxwmgbtmk}&
{\tt vfgxqdrkswgbpgz}&
{\tt vfgwpmmxhqgbkmc}&
{\tt svkfgxlujlwmgbx}\\
{\tt vwtsrlfgxhgbije}&
{\tt vpdtwfgkfkxgbkp}&
{\tt zvyfgsdwxjgbzdn}&
{\tt kvdsovwfgxgbyhf}\\
{\tt vmnfgmxxwigbcc}&
{\tt qwvqfgfwxxfegbg}&
{\tt wzcagczvfcaxgab$^{(1)}$}& \\
      \bottomrule
      
    \end{tabular}
  
   \label{tab:seq}

\end{document}